\documentclass[%
 aip,
 jcp,%
 amsmath,amssymb,
 reprint,%
]{revtex4-1}

\usepackage{graphicx}
\usepackage{dcolumn}
\usepackage{bm}

\usepackage{lipsum}

\newsavebox{\fmbox}

\begin{document}


\title{Non-linear eigensolver-based alternative to traditional SCF methods}

\author{B. Gavin}
\author{E. Polizzi}
\affiliation{Department of Electrical and Computer Engineering, 100 Natural Resources Road, Marcus 201,
 University of Massachusetts, Amherst, USA}

\date{\today}

\begin{abstract}

The self-consistent procedure in electronic structure calculations is revisited using a highly efficient and robust 
algorithm for solving the non-linear eigenvector problem i.e. $H(\{\psi\})\psi=E\psi$.
This new scheme is derived from a generalization of the FEAST eigenvalue algorithm 
to account for the non-linearity of the Hamiltonian with the occupied eigenvectors. 
Using a series of numerical 
examples and the DFT-Kohn/Sham model, it will be shown that our approach can outperform the traditional SCF 
mixing-scheme techniques by providing a higher converge rate, convergence to the correct solution regardless 
of the choice of the initial guess, 
and a significant reduction of the eigenvalue solve time in simulations. 

\end{abstract}
\maketitle

\section{Introduction}

\begin{figure}[htbp]
\centering
\includegraphics[width=\linewidth]{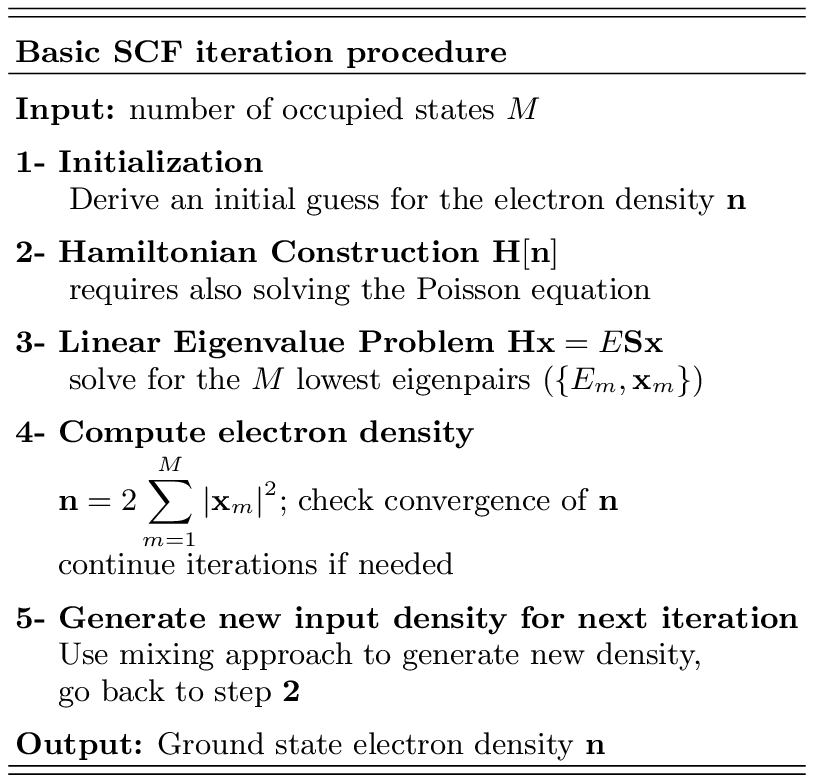}
\caption{\label{fig1} Basic self-consistent procedure for solving the (discretized) non-linear eigenvector problem
\mbox{${\bf H[n]x}={E}{\bf Sx}$} with $\bf H$ real symmetric and $\bf S$ symmetric positive definite (\mbox{$\bf S\neq I$} using
non-orthogonal basis functions), and obtaining the ground state electron density ${\bf n}=2\sum_m |{\bf x}_m|^2$ (with a factor $2$ for spin).
For the DFT/Kohn-Sham problem, the Hamiltonian $\bf H[n] = -\Delta + V_H[n]+V_{XC}[n]+V_{ext}$, 
is composed of  the Hartree potential $\bf V_H$ (solution of Poisson equation), the exchange-correlation potential 
$\bf V_{XC}$, and other external potential $\bf V_{ext}$ including the ionic potential.}
\end{figure}

Although first-principle calculations in general, and DFT in particular, have provided a practical 
(i.e. numerical tractable) 
path for solving the electronic structure problem, they have  introduced new numerical challenges on their own.
Within the single electron picture,  the Hamiltonian operator depends on the occupied eigenfunctions 
and the resulting eigenvalue problem becomes fully non-linear (i.e.  $H(\{\psi\})\psi=E\psi$).
In practice, this {\em non-linear eigenvector problem} is commonly addressed
using a self-consistent field method (SCF) wherein a series of linear eigenvalue problems (i.e.  $H\psi=E\psi$),
needs to be solved iteratively until convergence \cite{martin}.
Successfully reaching convergence by performing SCF iterations is of paramount importance to first-principle quantum chemistry
 and solid-state physics simulations software.
A typical self-consistent iteration procedure for the discretized DFT/Kohn-Sham problem  is represented in Figure \ref{fig1}.

Traditional SCF mixing methods
 employ successive approximation iterates of a fixed point mapping 
to generate the new input electron density at each cycle.  Examples of such methods include
 Newton-Broyden, or other Anderson technique,  and Pulay mixing techniques using direct inversion of the iterative subspace 
(DIIS) \cite{pulay,eyert96,grpulay,rohwedder,garza}.
Using DIIS, the new input electron density at iteration $k$, is generated from a linear combination of a series of previous trial densities
(i.e. ${\{{\bf n}^{(k-1)},{\bf n}^{(k-2)},{\bf n}^{(k-3)}, \dots \}}$) by minimizing the successive residual errors between input 
and output densities.

Three main numerical difficulties arise from standard SCF procedures: 
(i) the linear eigenvalue problem  needs to be solved repeatedly a large number of times;
(ii) the robustness of the  self-consistent iterations is very sensitive to the choice of the initial guess; and
(iii) there is, as of yet, no robust and efficient general-purpose
iterative numerical scheme for addressing the non-linear coupling with guaranteed convergence.
Although much progress has been made to improve the converge rate of SCF techniques \cite{kudin,host,xhu,ywang,kbaarman,kbaarman2},
 the iterations can still be found to converge very slowly or unreliably \cite{yang09}. 

In this paper, we present an efficient alternative to traditional SCF iteration techniques that is ideally suited to 
address the aforementioned numerical difficulties when 
solving the discretized non-linear eigenvector problem 
\mbox{${\bf H[n]x}={E}{\bf Sx}$}.
The SCF problem is fully revisited using a 
general-purpose numerical strategy derived from a modification of the FEAST eigenvalue algorithm \cite{p2009}.
 We show that the new approach, named NLFEAST (for Non-Linear FEAST), takes naturally advantage of the intrisic 
subspace iterations procedure of the FEAST algorithm to achieve global convergence, while 
the non-linearity is only addressed at a level of a reduced system (which can be solved even approximately 
using any SCF procedures).

The outline of this paper is as follows: In section \ref{sec:feast} we
briefly summarize the numerical steps and the main properties of the
FEAST algorithm presented in Ref. \onlinecite{p2009} for solving the linear
eigenvalue problem. In section \ref{sec:nlfeast} we present the generalization of FEAST 
for solving the non-linear eigenvector problem.
Numerical results and capabilities of the new NLFEAST approach are then presented and discussed 
in Section \ref{sec:result}.

\section{The FEAST Algorithm}\label{sec:feast}

Within the SCF procedure in Figure \ref{fig1}, solving the 
linear and symmetric eigenvalue problem
 at a given iteration step becomes the  most time-consuming part of the electronic structure calculations.
In such computations, one can identify two main challenges: (i) discretization techniques 
that accommodate atomistic systems with a high level of accuracy 
can lead to large size system matrices {\bf \{H, S\}}, and (ii) the number of eigenpairs needed to 
compute the electron density  is proportional to the number of atoms in the system.
In order to characterize large-scale nanostructures and complex systems of current technological interest,
 many thousands of eigenpairs are indeed needed.  In this regard, progress in large-scale electronic 
structure calculations can be tied together 
with advances in numerical algorithms for addressing the eigenvalue problem. 

\begin{figure}[htbp]
\centering
\includegraphics[width=\linewidth]{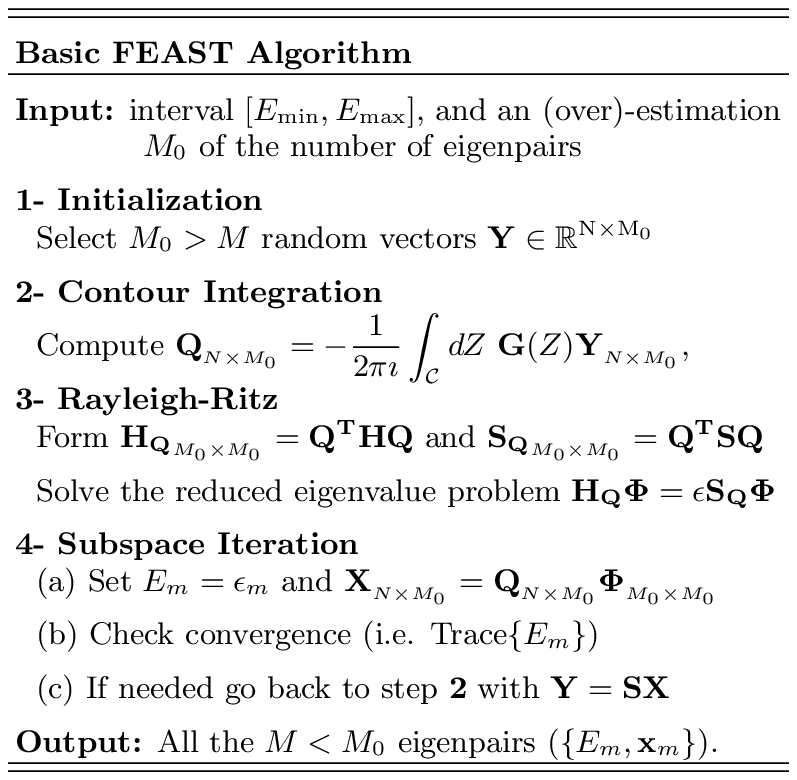}
\caption{\label{fig2} Basic FEAST procedure for solving the generalized eigenvalue problem
\mbox{${\bf Hx}={E}{\bf Sx}$} of size $N$ with $\bf H$ real symmetric and $\bf S$ symmetric positive definite (spd), 
and obtaining all the $M$ eigenpairs within a given interval $[E_{\rm min},E_{\rm max}]$.
The density matrix appears implicitly in Step-2, using the complex contour integration of the Green's function 
\mbox{${{\bf G}({Z})}= ({Z}{\bf S}-{\bf H})^{-1}$}. In practice, the vectors $\bf Q$ in Step-2 can be computed using a high-order numerical
 integration such as Gauss-Legendre quadrature, where only a small number of linear systems, 
$(Z_e{\bf S}-{\bf H}){\bf Q_{e}=Y}$,
 need to be solved, one for each of
 a number of specific Gauss nodes $Z_e$ (associated with the weights $\omega_e$) along a complex contour ${\cal C}$,  i.e. 
${\bf Q}:=\sum_{Z_e\in {\cal C}}\omega_e{\bf Q_{e}}.$}
\end{figure}

The recent FEAST algorithm \cite{p2009},
which uniquely combines accuracy, robustness, high-performance and (linear) parallel scalability, is
ideally suited for addressing the electronic structure problem.
FEAST is a general algorithm
that can be used for solving the linear 
 eigenvalue problem
to obtain all the eigenvalues and eigenvectors within a given search interval  (e.g. $[E_{\rm min},E_{\rm max}]$).
FEAST's main computational tasks consist of solving a small number of independent linear systems with multiple right-hand 
sides along a complex contour and one reduced dense eigenvalue problem that is orders of magnitude smaller than
 the original one (the size of this reduced problem is of the order of the number of eigenpairs inside the search interval).
A full and detailed accounting of the implementation of the FEAST algorithm is given in Ref. \onlinecite{p2009}, and the basic
 procedure is briefly summarized in Figure \ref{fig2}. 

Worthy of particular note are Steps 2 and 4(c) which, combined, generate a subspace {\bf Q} that spans
the eigenvectors solutions in the user-defined interval.  
This is done by multiplying a given approximate eigenvector solution at iteration $k$ by the density matrix $\rho$ i.e.
\begin{equation}
{\bf Q}_{_{N\times M_0}}^{k+1}=\rho {\bf SX}_{_{N\times M_0}}^k,
\label{eq:rho}
\end{equation}
where the new subspace ${\bf Q}^{k+1}$ is then obtained at the FEAST iteration $k+1$. 
Denoting by \mbox{${\bf X}_{_{N\times M}}=\{{\bf x}_1,{\bf x}_2,\dots,{\bf x}_M\}$} 
 the $M$ eigenvectors within the search interval, we note that the density matrix is formally 
given by $\rho={\bf X}{\bf X^T}$.
 In practice, the spectral projection (\ref{eq:rho})
is conveniently obtained by numerical integration of a set of solutions of independent linear systems defined 
along a complex contour.
It can be shown that FEAST consists of a reformulation of the subspace iteration technique 
for spectral projectors \cite{pp12},
which leads to an extremely robust and accurate numerical procedure.
By this means the linear FEAST algorithm achieves very high convergence rate of the eigenvectors subspace, and allows 
the non-linear extension of FEAST (i.e. NLFEAST) to achieve the kind of performance that will be demonstrated in this paper.

The FEAST algorithm holds  all the following important intrinsic properties:
\begin{itemize}
\itemsep 1pt
\parskip 1pt
\item using a high-order Gauss-Legendre quadrature, $8$ to $16$ contour 
points suffice for FEAST to consistently converge in $\sim $3 iterations to obtain up to thousands of eigenpairs with machine accuracy; 
\item all multiplicities are naturally captured; 
\item no (explicit) orthogonalization procedure is required; 
\item pre-computed subspace can be reused as suitable initial guess (i.e. Step-1 in Figure \ref{fig2}) 
for solving a series of eigenvalue problems that are close one another.
One can ideally take advantage of this feature for addressing, in particular: (i) Step-3 of the SCF procedure in Figure \ref{fig1} 
along the iterations, (ii) bandstructure calculations \cite{p2009} along the $k$ space, or 
(iii) time-dependent propagation using spectral decomposition along the time-domain \cite{cp2010}; 
\item the inner linear system at each contour point can be solved using either direct or iterative methods; 
\item the algorithm can be extended efficiently for solving non-Hermitian problems (e.g. for performing complex bandstructure calculations \cite{laux});
\item efficient parallel implementations can be naturally addressed at three different levels:
(i) many search intervals can be run independently (no overlap), (ii) each linear system can be solved 
independently  along the complex contour (e.g. simultaneously on different compute nodes), and  (iii) 
the linear systems can be solved in parallel (the multiple right sides can be parallelized as well).
Since the search intervals can be arbitrarily narrowed 
within a parallel environment (i.e. increasing the ratio $N/M_0$), 
the algorithm complexity is directly dependent on solving 
a single linear system of size $N$ with $M_0$ right-hand-sides (i.e. $(Z_e{\bf S}-{\bf H}){\bf Q_{e}=Y}$ in Figure 
\ref{fig2}). 
\end{itemize}

\section{Direct solution of the non-linear eigenvector problem}\label{sec:nlfeast}

Here, we propose to further extend the capabilities of the FEAST algorithm in order to solve nonlinear eigenvector problems 
of the form 
\begin{equation}
 {\bf H[n]}{\bf x}_i=E_i {\bf Sx}_i,\quad {\bf n}=2\sum_i |{\bf x}_i|^2,
\label{eq:nl}
\end{equation}
so that it may be used as a direct and efficient alternative to SCF methods
conventionally based on  a ``Schr\"odinger-Poisson'' iterative procedure with density mixing schemes. 

Recently Yang et al. \cite{yang06} have proposed an alternative approach to 
 traditional SCF iterations using a direct 
constrained minimization (DCM) algorithm for solving the nonlinear problem (\ref{eq:nl}). 
DCM consists of  constructing and updating new search directions for the eigenvector subspace by solving a much smaller optimization 
problem which takes the form of a reduced nonlinear eigenvector problem. These results have motivated this current work, since
a reduced eigenvalue problem can also be obtained from the FEAST algorithm, which also
provides an optimal framework for performing subspace iterations.
Indeed, once the search subspace ${\bf Q}$ is obtained
by the contour integration in Step-2 of Figure \ref{fig2} (from a given Hamiltonian $\bf H[n]$),
 the resulting reduced  problem in Step-3 can be expressed in a non-linear form i.e. 
 \begin{equation}
  {\bf H_Q[n]\Phi}=\epsilon {\bf S_Q\Phi},  \quad \mbox{ with } \quad  {\bf n}=2\sum_i |{\bf Q\Phi}_i|^2,
 \label{eq_nlr}
 \end{equation}
 with $\bf H_Q[n]=Q^T H[n]Q$.  Using a DFT/Kohn-Sham  model, $\bf H_Q$  accounts for the projection of the non-linear 
Hartree and exchange-correlation terms onto the working search subspace ${\bf Q}$.
One of the primary difference between linear
 FEAST and the proposed non-linear FEAST algorithm, named NLFEAST, then lies in Step-3
 with the formation and solution of the reduced non-linear eigenvector problem.
This latter, in turn, can be solved using the traditional SCF procedure presented in Figure \ref{fig1}.
Since the non-linearity appears at the level of the reduced system alone,  
more robust non-linear schemes that would have been too expensive to use on the large size original system (\ref{eq:nl}),  
can potentially be considered for addressing the reduced system (\ref{eq_nlr}) (e.g.  specific Newton-Raphson method \cite{Jerome}).

As in the linear case, once the solution of the reduced system is obtained, 
 the subspace is updated and the FEAST algorithm can come back on track and perform subspace iterations up until convergence.
As it will be demonstrated in our experimental results in Section \ref{sec:result}, however, 
the algorithm essentially converges 
faster when the size of the subspace is increased with each FEAST subspace iteration.
These results suggest that the size of the search subspace should not be made static
 for the non-linear problem.
This then constitutes 
a major and innovative 
difference with the original FEAST algorithm that aims at 
improving the robustness of the subspace iteration technique.
The Raleigh-Ritz procedure (i.e. Step-3) then uses a new search subspace $\bf \hat{Q}$ which 
consists not only of the subspace most recently
produced by solution of the contour integration at step $k$, i.e. $ {\bf Q}^{(k)}$,
 but also of all (or some) of the subspaces generated in previous iterations such that ${\bf \hat{Q}}=\{{\bf Q}^{(k)}, {\bf Q}^{(k-1)},{\bf Q}^{(k-2)},\dots\}$. 
The subspace provided by the contour integration can be appended to  ${\bf \hat{Q}}$ indefinitely until convergence, or 
this can be done such that only a finite number (typically $3$ or $4$) of the most recently generated subspaces are 
retained. Interestingly, this procedure is reminiscent of the one used in DIIS
where new electron densities are constructed from a subspace consisting of previous generated electron densities and density
residuals. However, traditional mixing schemes act on the electron density alone, and it is not always guaranteed 
that the density can be expressed as a linear combination of previously generated densities.
In contrast, the proposed approach is expected to be much more robust by essence since
 a larger and refined search subspace containing all the eigenvectors of interests,
is very likely to span the solution of the non-linear problem.

\begin{figure}[htbp]
\centering
\includegraphics[width=\linewidth]{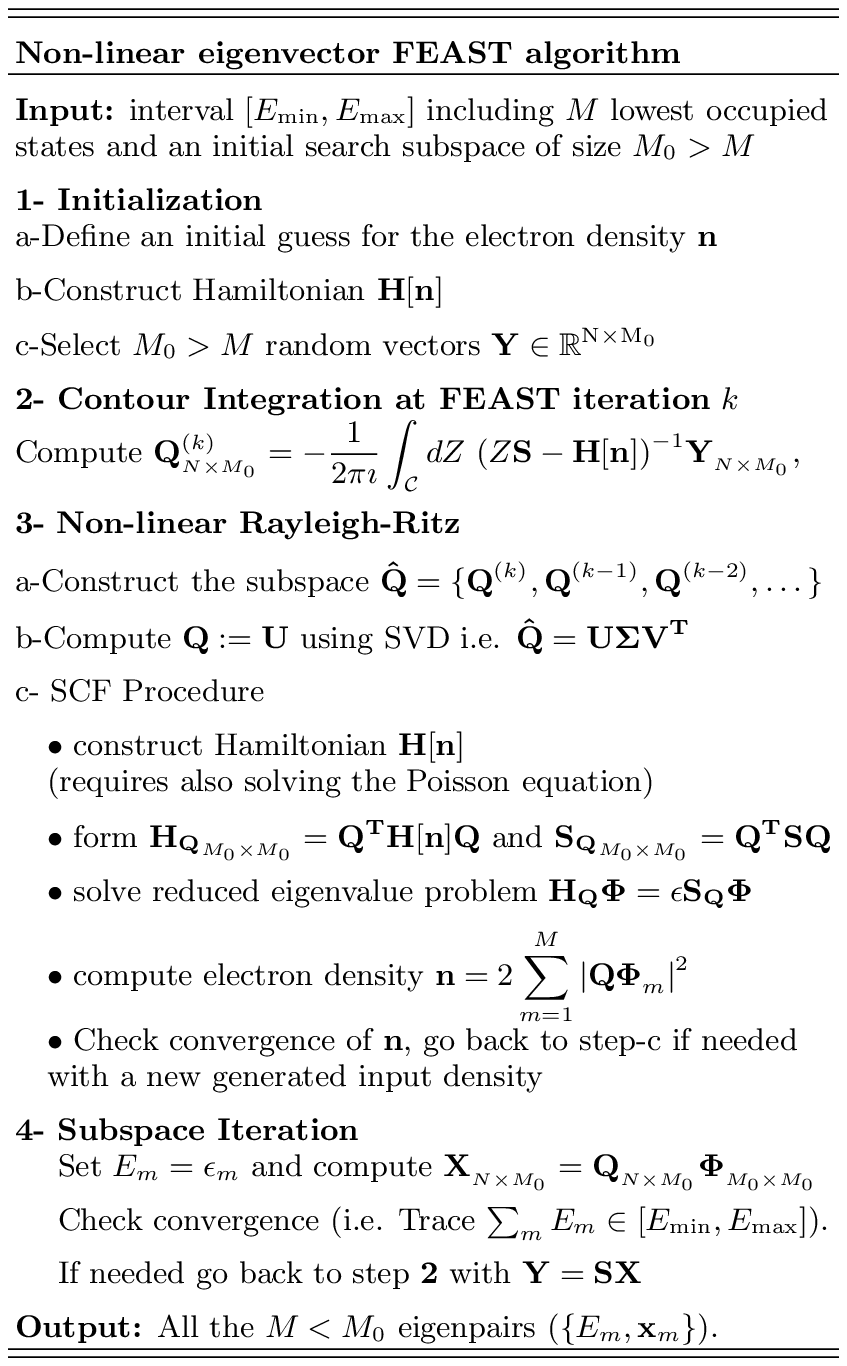}
\caption{\label{fig3} FEAST procedure for solving the non-linear eigenvector problem
\mbox{${\bf H[n]x}={E}{\bf Sx}$} of size $N$ with $\bf H$ real symmetric and $\bf S$ 
symmetric positive definite (spd), and obtaining all the $M$ lowest occupied states within a given interval 
$[E_{\rm min},E_{\rm max}]$ and hence the ground state electron density. Here, Step 3-c of the algorithm is making use of 
a traditional SCF procedure for solving the non-linear reduced system. We note that in Step-3a,
 the column vectors of the newly generated subspace are appended to the matrix of the old subspace, increasing the total subspace size
 by $M_0$. This can be done such that only a finite number of the most recently generated subspaces are retained, or the subspace size 
can be increased until convergence.  Additionally, a singular value decomposition (SVD) is performed on $\bf \hat{Q}$ (i.e. Step 3-b), 
such that only the left singular vectors $\bf U$ are used in the Rayleigh-Ritz procedure.}
\end{figure}

The resulting NLFEAST algorithm is given in Figure \ref{fig3}.
We note that a singular value decomposition (SVD) also needs to be performed on the search subspace $\bf \hat{Q}$ (i.e. Step 3-b),
 where only the left-singular vectors are computed and retained.   In contrast to the linear FEAST algorithm, 
one does not want to truncate the size of the search subspace  for constructing 
a positive definite reduced matrix $\bf S_Q$. This latter is by definition rank deficient and the orthogonalization
of the subspace cannot then be avoided (the SVD step can also be replaced by a QR decomposition, although the singular
 values could potentially provide additional information about the convergence rate \cite{pp12}).
In our experiments thus far this step has proven to be computationally inexpensive relative to 
other parts of the algorithm, 
and even so, it can be shown that the SVD of $\bf{\hat{Q}}$ can be updated at each successive FEAST 
iteration with complexity similar to solving 
a single reduced eigenvalue problem in Step-3c \cite{sameh}.
Finally, like other schemes for solving nonlinear eigenvector equations, NLFEAST requires an initial guess for the density ${\bf n}$
 (Step-1a of Figure \ref{fig3}). However, the quality of the initial guess is not important as far as achieving
convergence is concerned, as it will be shown in Section \ref{sec:result}.

In summary, NLFEAST can be seen as an inversion of the usual process of solving the nonlinear eigenvector problem (\ref{eq:nl}).
Using SCF-DIIS, one iteration corresponds to a traditional SCF iteration as presented in Figure \ref{fig1} which requires, in particular, 
solving a large linear eigenvalue problem as well as the calculation of the Hartree potential by solving Poisson's equation. 
This scheme presents then two iterative procedures: one outer associated with the SCF, and one inner due to the eigenvalue solver 
(using a ``black-box'' eigenvalue solver, those inner iterations -intrinsic to eigenvalue algorithms- are most often hidden to the user).
As discussed earlier, this linear eigenvalue problem can ideally be solved using the linear FEAST algorithm (in Figure \ref{fig2}) 
which converges in very few iterations of contour integration (typically $\sim 3$). Using NLFEAST, however, those two iterative procedures happen 
now in reverse order. The outer iterations are intrinsic to the FEAST algorithm (i.e. subspace iterations), while the inner iterations are used for solving 
the non-linear reduced system (\ref{eq_nlr}). In addition to offering a more robust mathematical 
approach, NLFEAST is expected to considerably reduce the eigenvalue solve time since: 
(i) the reduced non-linear problem (\ref{eq_nlr})
is orders of magnitude smaller in size as compared to the original one (\ref{eq:nl}), and (ii)
 the reduced problem does not need to be solved very accurately 
 to guarantee the  convergence of the outer-iterations.

\section{Numerical Experiments and Capabilities}\label{sec:result}

In this section, we propose to demonstrate the numerical efficiency 
of the proposed NLFEAST algorithm for the non-linear eigenvector problem (\ref{eq:nl}).
 DFT/Kohn-Sham/LDA calculations 
have been performed on various molecules using our in-house all-electron simulation framework that uses a real-space cubic finite element discretization.
Some details of our modeling and real-space discretization setup have been provided in Ref. \onlinecite{lzp12}. 
The Pulay-DIIS technique has been used for
performing the traditional SCF approach defined in Figure \ref{fig1}, as well as for solving
the FEAST non-linear reduced system in Step-3c of Figure \ref{fig3}. 
Both DIIS procedures make use of a subspace composed of 
five successive generations of the electron densities, but the maximum number of DIIS iterations for solving the reduced system in 
NLFEAST has been fixed to three (i.e. the non-linear reduced system is then solved only approximately). For most of the examples, indeed, increasing 
the number of the inner iterations further has had no effect on the overall NLFEAST convergence rate.

\subsection{Performance comparisons}\label{sec:resultA}

\begin{figure}[htbp]
\centering
\includegraphics[height=\linewidth,angle=-90]{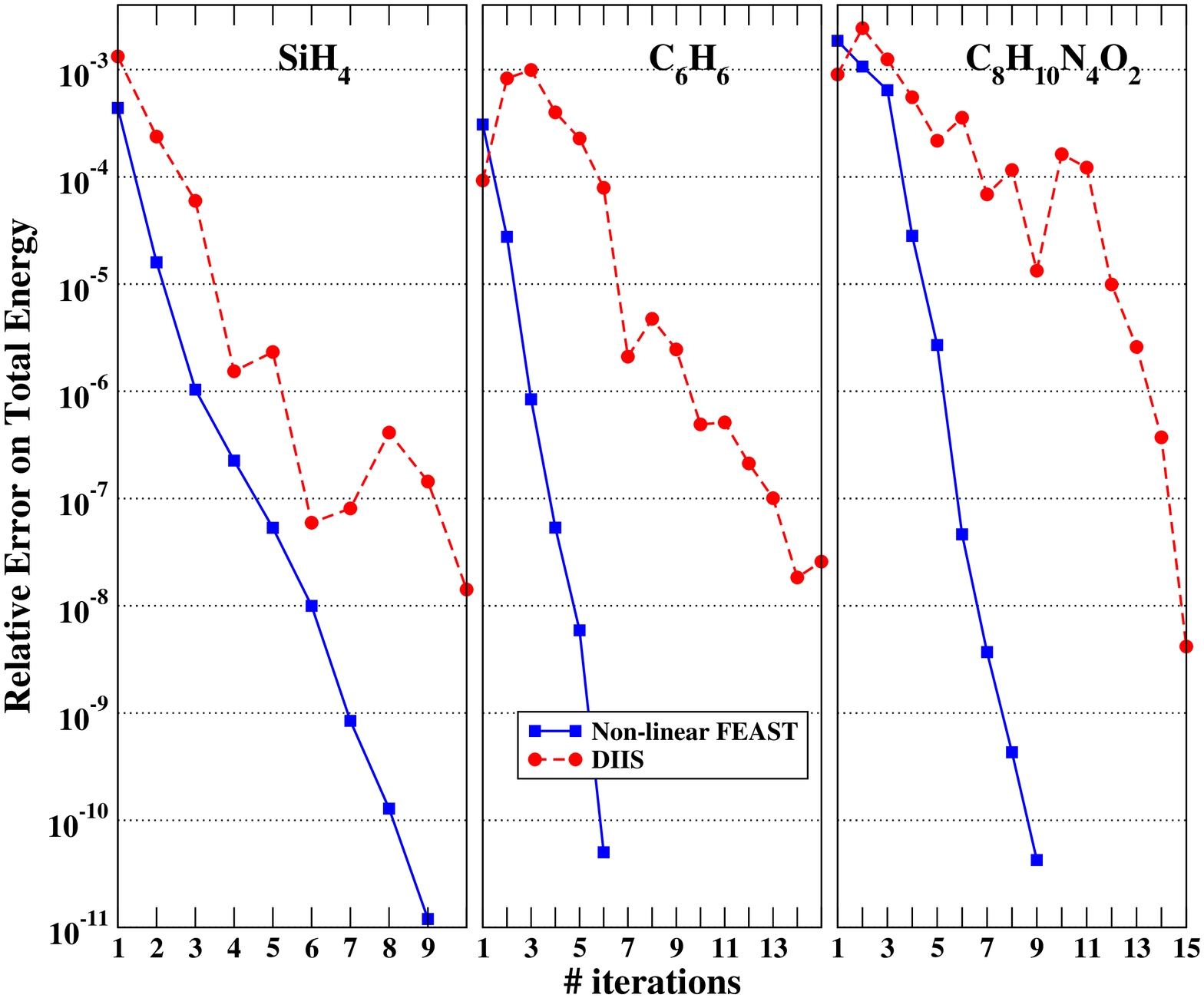}
\caption{\label{fig4} Results of numerical experiments comparing the performance of our algorithm to that of DIIS for the Silane $\rm SiH_4$, 
Benzene $\rm C_6H_6$ and caffeine $\rm C_8H_{10}N_4O_2$ molecules.  The relative error on the total energy is used here as the measure of
convergence. 
The convergence criteria for NLFEAST has been set to $10^{-10}$ while DIIS-SCF was stopped after a 
 maximum of $10$ iterations for Silane and $15$ iterations for both Benzene and Caffeine. The meaning of outer-iterations is different for both
algorithms.}
\end{figure}

In all our experiments conducted thus far, including various molecules from $\rm H_2$ to $\rm C_{60}$,
 the NLFEAST algorithm  has outperformed SCF-DIIS both in terms of convergence rate 
and execution time. Three representative examples are shown in Figure \ref{fig4} for the Silane, Benzene and Caffeine molecules.
The relative error on the total energy at each iteration is used as the measure of convergence for both approaches,
 as this is the most directly comparable measure between DIIS-SCF and NLFEAST.
As discussed in Section \ref{sec:nlfeast},  the meaning of the outer-iterations is different for both algorithm since, for NLFEAST, it directly 
represents the number of contour integrations and non-linear reduced systems.

Since DIIS-SCF also takes advantage of the linear FEAST algorithm, it is possible to have a direct comparison 
between the main numerical operation counts of DIIS-SCF and NLFEAST. Table \ref{tab_compcost} summarizes the number 
of FEAST contour integrations and Poisson system solves
 needed by both approaches for obtaining the same convergence 
accuracy (i.e. $\sim 10^{-8}$) for the three molecules.
\begin{table}[htbp]
\begin{small}
\begin{center}
\begin{tabular}{lcc|cc|cc} \hline\hline
&\multicolumn{2}{c}{$\rm SiH_4$} & \multicolumn{2}{c}{$\rm C_6H_6$} & \multicolumn{2}{c}{$\rm C_8H_{10}N_4O_2$} \\ 
 & {\footnotesize NLFEAST} & {\footnotesize SCF}  & {\footnotesize NLFEAST} & 
{\footnotesize SCF}  & {\footnotesize NLFEAST} & {\footnotesize SCF} \\ \hline
\# Contour & 7 & 45 & 6 & 60 & 8 & 60 \\
\# Poisson & 56 & 15 & 48 & 20 & 64 & 20\\
\hline\hline
\end{tabular}
\end{center}
\caption{\label{tab_compcost} Comparison of the number of contour integrations and Poisson system solves required to 
reach convergence (set at $\sim 10^{-8}$) using NLFEAST and SCF-DIIS and the DFT/Kohn-Sham/LDA model for 
the three molecules  
$\rm SiH_4$, $\rm C_6H_6$ and $\rm C_8H_{10}N_4O_2$. The number of additional iterations needed to construct 
the initial subspace (not reported in Figure \ref{fig4}) has also been taken into consideration 
(i.e. one additional contour for NLFEAST and five DIIS iterations for SCF-DIIS). 
With enough parallelism for FEAST, 
the cost of each contour integral can be straightforwardly reduced to that of solving a single complex linear system, 
whereas the solution of Poisson's equation consists of solving a single real-valued linear system.  
}
\end{small}
\end{table} 
From these results, one can conclude that
SCF-DIIS requires $6\times$ to $10\times$ more 
 contour integrations than NLFEAST for obtaining the same level of accuracy. 
On the other hand, NLFEAST requires the solution 
of $2.5\times$ to $4\times$ more Poisson equations as does SCF-DIIS.
 Using the DFT/Kohn-Sham/ALDA model, in particular, 
the Poisson solve operations are much less expensive 
than factorizing and solving the complex linear systems that arise at each contour integration.
Considering the operation counts in Table \ref{tab_compcost},
 the computational cost of NLFEAST is therefore expected to generally 
be much less than that of SCF-DIIS for the same problem. 
While comparing NLFEAST and SCF-DIIS, however, it is important to point out 
possible issues regarding the cost of the NLFEAST inner iterations for physical models different than pure DFT/LDA 
(such as Hartree-Fock) and where the Hamiltonian construction is more involved than solving a single Poisson equation.
One solution consists of keeping on cutting even further the number of inner iterations as long as it does not 
affect the global convergence.

Finally, the rest of the numerical operations involved in NLFEAST, such as solving the reduced eigenvalue problem or performing 
the SVD in Step-3c 
of Figure \ref{fig3}, bring no significant computational overhead provided the ratio $N/M_0$ stays relatively 
large for a given search interval. 
In practice, it is also reasonable to use an original subspace size $M_0$ that is typically no larger than 
$\sim 500$,  if we assume that the $\bf \hat{Q}$ subspace keeps increasing indefinitely and that convergence
 is reached in less than $10$ iterations. Since it is recommended for FEAST to use $M_0\sim1.5\times$ the number of eigenstates \cite{p2009},
the basic NLFEAST scheme is expected to provide good performances up to $\sim 350$ states (i.e. $700$ electrons with a factor 2 for spin).
It should be noted that the case of a smaller $N/M_0$ ratio with a relatively large $M_0$, can lead to degradation of performances 
and memory limitations only if the algorithm runs sequentially.
The FEAST algorithm can indeed be readily parallelized and the applicability of NLFEAST scheme for addressing large-scale molecular 
systems will be discussed in Section \ref{sec:discuss}.

Table \ref{tab_summary} summarizes the convergence results we have obtained using NLFEAST for a sample of molecules ranging 
in size from $\rm H_2$ to $\rm C_{60}$, with results on the total energy which are compared with the NWChem software \cite{nwchem}.
Although we did no attempt to optimize or refine our cubic finite element discretization,
the total energy results are found to be in good agreement with NWChem which is making use of completely different numerical approaches
 and convergence criteria (for these reasons also, a direct and useful comparison of the convergence rate cannot be achieved here).
It is also important to note that in the case of the non-linear eigenvector problem (\ref{eq:nl}), the relative error on the total energy 
is not a natural criteria for measuring the convergence.
For NLFEAST, the criteria is instead defined in terms of the error on the trace of the eigenvalues. 
From our numerical experiments, once this convergence criteria is satisfied, it also provides very low non-linear 
residual. 
For the simulations reported in Table  \ref{tab_summary}, we have also used:  
(i) a  $\bf \hat{Q}$ search subspace that keeps increasing indefinitely until convergence, (ii) $16$ nodes for the Gauss quadrature 
along the complex contour, (iii) a conventional initial guess 
as the starting point for the electron density (e.g. the all-electron result for the single atoms).
 In the following subsections, the robustness of NLFEAST will be discussed and demonstrated further by 
separately addressing each one of those points. 

\begin{table}[htbp]
\begin{center}
\begin{tabular}{lccccc} \hline\hline
& \#electrons & \#iterations &  $\rm E_{tot}$(eV)   & NWChem \\ \hline 
$\rm H_2$ &      2         &  5   & -30.962 & -30.959\\
$\rm CH_4$ &    10         & 5  & -1091.45& -1091.69\\
$\rm H_2O$ &    10         & 7  & -2065.24  & -2065.48 \\
$\rm CO$ &      14         & 6 &  -3060.12 & -3060.45\\
$\rm SiH_4$ &   18         &  9 &  -7907.42& -7909.78 \\
$\rm Na_2$ &    22         &  8  &   -8785.57 & -8786.61 \\
$\rm C_6H_6$ &  42         &    6 &  -6262.31 & -6263.65\\
$\rm C_8H_{10}N_4O_2$ & 102 &  9 &  -18364.5 & -18365.3 \\
$\rm C_{60}$         & 360&  8 &   -61676.7 & -61673.7\\
\hline\hline
\end{tabular}
\end{center}
\caption{\label{tab_summary} Convergence rate results for NLFEAST for various molecules  
(the molecular geometries are obtained from the experimental data \cite{cccbdb}). 
The convergence criteria is satisfied for NLFEAST when the relative error on trace is below $10^{-10}$, 
which also provides very low non-linear relative residual  
$\max_m(||{\bf H}({\bf x}_m){\bf x}_m-E_m{\bf Sx}_m||/||{\bf H}({\bf x}_m){\bf x}_m||)$, 
typically here below $10^{-8}$. 
The total energy results using our cubic real-space FEM code are in good agreement
 with those obtained by NWChem \cite{nwchem} using the $\rm cc-pvqz$ basis from $\rm H_2$ to $\rm C_6H_6$,
 and the $\rm 6-311g^*$  basis for $\rm C_8H_{10}N_4O_2$  and $\rm C_{60}$.}
\end{table}

\subsection{Convergence Rate and {\bf Q} subspace size}\label{sec:resultB}

\begin{figure}[htbp]
\centering
\includegraphics[height=\linewidth,angle=-90]{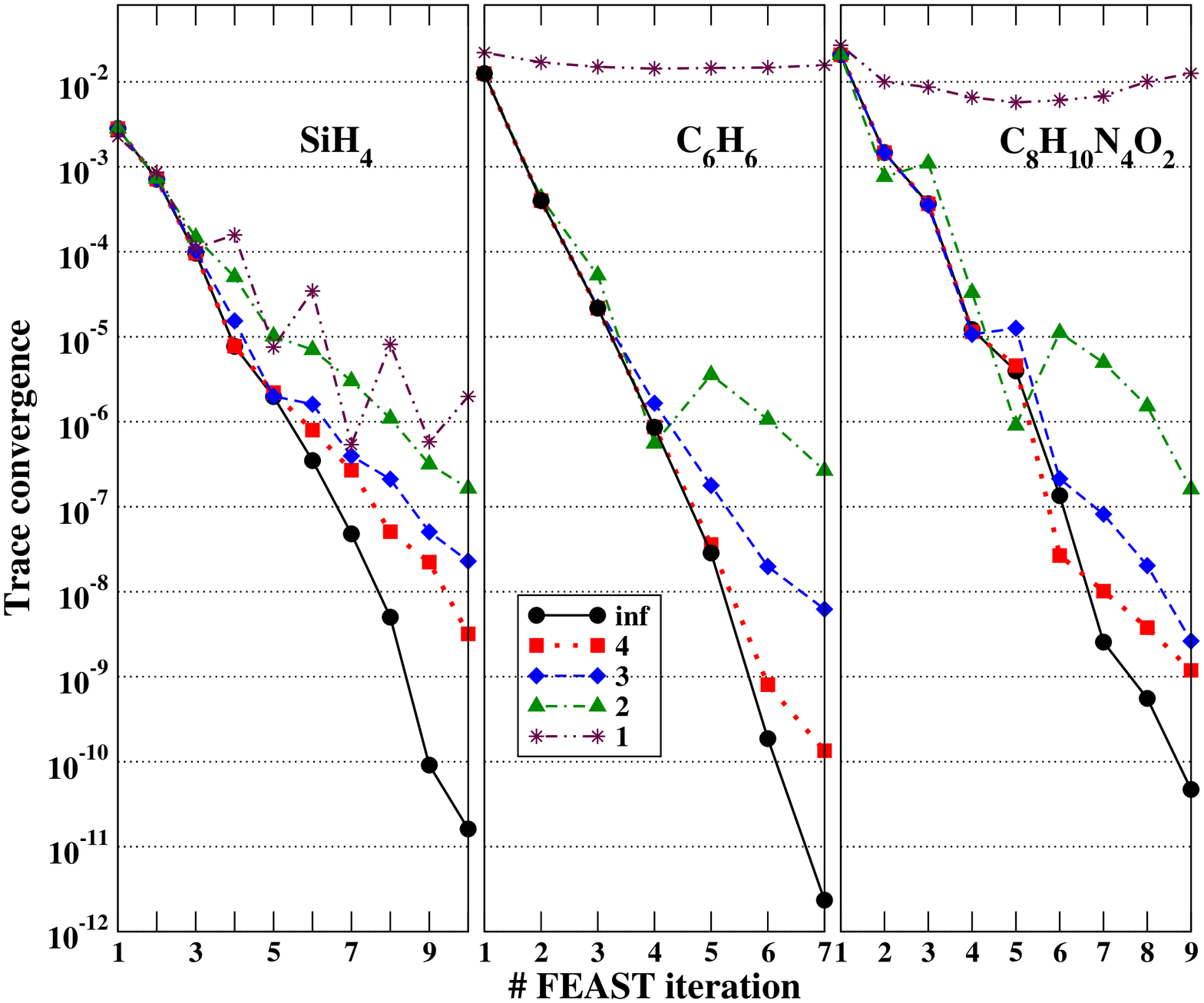}
\caption{\label{fig5} Numerical experiments demonstrating the role of the number of most recent retained ${\bf Q}^{(k)}$ subspaces for 
constructing the search subspace ${\bf \hat{Q}}=\{{\bf Q}^{(k)}, {\bf Q}^{(k-1)},{\bf Q}^{(k-2)},\dots\}$ used by NLFEAST. The convergence rate depends on 
this number that here varies  between one and ``infinity'' (i.e. where the search subspace keeps increasing until convergence is reached).}
\end{figure}

The size of the search subspace $\bf \hat{Q}$ (in step-3a of Figure \ref{fig3}) may be limited by only retaining a finite number 
of the most recent subspaces, or it may be extended indefinitely until convergence.
For the numerical experiments presented in Figure \ref{fig4} and Table \ref{tab_summary}, the rate of convergence is expected to reach
 a maximum because all generated subspaces have been retained. 
Figure \ref{fig5} shows the results of several numerical experiments that were performed for our selected three molecules using
different sizes for the search subspace  ${\bf \hat{Q}}$. 
From these results, one observe that the rate of convergence of NLFEAST is directly related to the number of subspaces that are 
retained, but a high rate of convergence can still be obtained by retaining only three to four of the most recent subspaces.
By retaining a limited number of subspaces,  
it becomes also possible to increase the size of $M_0$ to consider a larger number of electrons for a given search interval 
without resorting yet to explicit parallelism. 
In the case of Benzene and Caffeine, one also notes that the algorithm does not appear to converge at all when only a single ${\bf Q}^{(k)}$ subspace 
is used, this subspace being the one that was most recently generated. Although this is not necessarily typical and the solution may eventually
converge using a more efficient approach for solving the non-linear reduced problem (e.g. 
using a larger number of inner iterations for the FEAST-DIIS problem which has been fixed to three in our simulations), 
it highlights the importance of extending the search subspace size for the success of this algorithm.

\subsection{Convergence Rate and Contour Integration Accuracy}

At each iteration of NLFEAST, the approximate subspace solution is improved through multiplication by the density matrix
of the most current Hamiltonian (see equation \ref{eq:rho}).  
This step, Step 2 in Figure \ref{fig3}, 
is accomplished by performing a numerical contour integration of the Green's function multiplied by the approximate solution. In practice, 
a $n$-point Gauss quadrature can efficiently be used here, which involves
summing the solutions of $n$ separate linear systems.   
\begin{figure}[htbp]
\centering
\includegraphics[height=\linewidth,angle=-90]{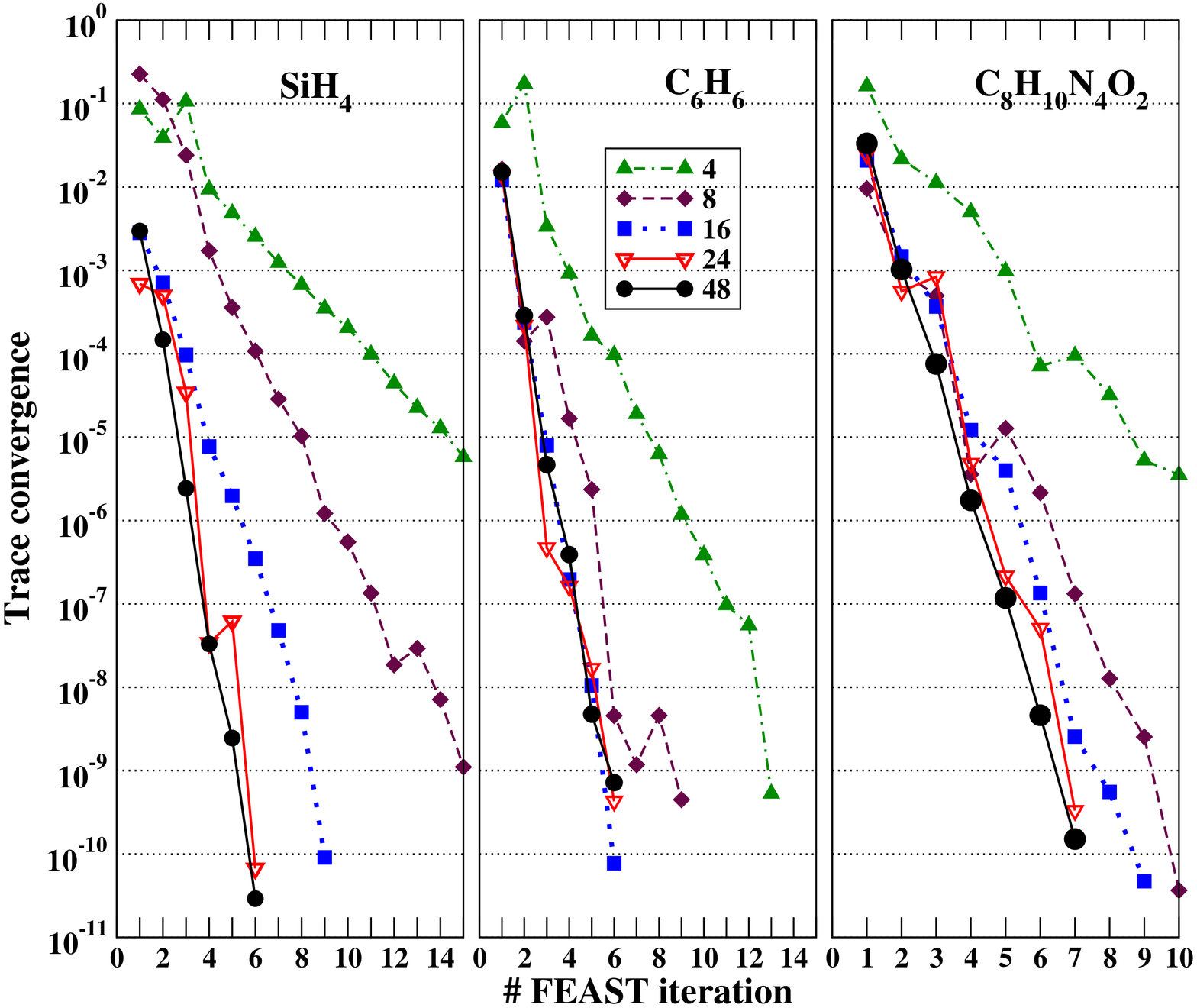}
\caption{\label{fig6} Experiments demonstrating the effect on convergence of the accuracy of the contour integration in NLFEAST. 
Different curves represent the convergence
for different numbers of Gauss points used in the numerical contour integration (Step 2 in Figure \ref{fig3}).
For all these results,  the search subspace $\bf \hat{Q}$ keeps increasing indefinitely until convergence.}
\end{figure}
For all the results presented so far, 
$16$ Gauss contour points were used to perform the quadrature. Figure \ref{fig6} presents 
a new set of convergence results
 for our three selected molecules, 
while considering the variation of the number of Gauss points from as low as $4$ to as high as $48$. 
Although increasing the number of Gauss points can help 
to improve the convergence rate, the effect here is not quite as dramatic as the effect of increasing the subspace 
size as discussed  previously. We note here very similar convergence behaviors between NLFEAST and 
the FEAST algorithm for the linear problem. This latter admits a mathematical convergence proof
that shows that the actual convergence rate depends on both the accuracy of the contour
 integration and the size of the search subspace \cite{pp12}.

\subsection{Robustness and initial guess}\label{sec:resultC}

\begin{figure}[htbp]
\centering
\includegraphics[height=\linewidth,angle=-90]{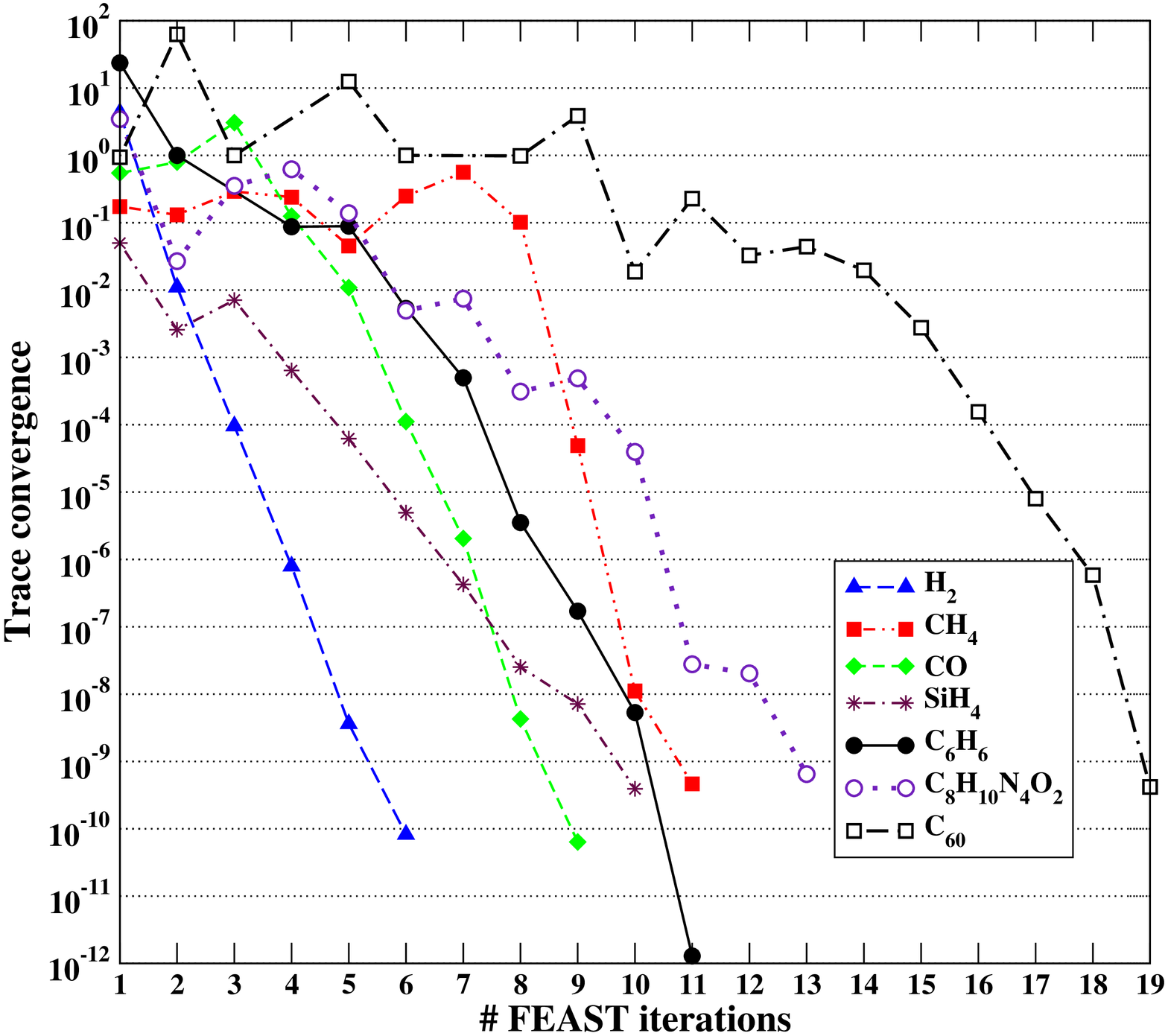}
\caption{\label{fig7} Results of numerical experiments where the initial guess for the electron density was set to zero,  ${\bf n}=0$. The maximum number of
subspace to form $\bf \hat{Q}$ has been set equal to $10$ for all molecules but
$\rm C_8H_{10}N_4O_2$  and $\rm C_{60}$ where it was set equal to $4$. The convergence is reached for an error on the trace smaller than $10^{-9}$}
\end{figure}

Like other schemes for solving nonlinear eigenvector equations, NLFEAST requires an initial guess for the density ${\bf n}$. 
Clearly, a good initial guess can provide faster convergence, but for NLFEAST
the quality of the initial guess is not important as far as achieving convergence is concerned. 
Unlike other means of performing SCF iterations, our algorithm is capable of achieving convergence even when given 
an extremely poor initial guess  including the extreme case of no initial guess at all (i.e.  ${\bf n}=0$). 
Figure \ref{fig7} shows the results of a number of numerical experiments wherein the initial guess 
for the electron density was set to zero. Each experiment eventually resulted in convergence.
The algorithm's performance when given such a poor initial guess is partly related to the size of the system; more electrons typically means 
that a larger number of iterations is required before convergence is reached. Performance also appears to depend on the particular molecule under
 consideration. In Figure \ref{fig7}, $\rm CH_4$ requires a larger number of iterations to reach a high 
 level of convergence,  despite being the second smallest system (in terms of number of electrons). 
It is likely that, for some molecules, the algorithm converges towards 
a local energy minimum before ultimately finding its way to the global minimum, which results in its progress being more delayed than it would 
be with a molecule where such a detour does not occur. 
The purpose of Figure \ref{fig7} is just to illustrate the robustness
of the algorithm by considering an extreme (academic) case. As shown in Table \ref{tab_summary}, a conventional initial guess has, so far,
 guaranteed low residual convergence in less than $10$ FEAST iterations for all the molecules that we have experimented with.


\section{Conclusion and future work}\label{sec:discuss}

A new eigensolver-based strategy, named NLFEAST, is derived
from  a generalization of the FEAST eigenvalue algorithm
for solving the non-linear eigenvector problem 
 i.e. $H(\{\psi\})\psi=E\psi$ such as the one arising from 
 the DFT/Kohn-Sham electronic structure model.
By providing  a fundamental and practical numerical solution for addressing the non-linearity with the occupied eigenvectors, 
NLFEAST
offers  a very efficient and robust alternative to the traditional self-consistent procedure using density-mixing schemes such as Pulay-DIIS. 
Several  numerical experiments using the DFT/Kohn-Sham/LDA model, have 
demonstrated the significant potential of NLFEAST to outperform 
the traditional SCF mixing techniques in terms of convergence rate, robustness, and numerical operation counts.
Strictly speaking, NLFEAST should not be considered as a direct competitor to current methods 
and it can rather be seen as an alternative formulation on how to handle the non-linearity.
Indeed, the resulting non-linear reduced system can, in turn, be addressed using any SCF procedures.
In our simulations, the reduced system did not need also to be solved 
accurately since the overall convergence  of our scheme benefits from the robustness 
of the FEAST subspace iterations (which also limits the number of inner SCF iterations).
This feature could potentially be further exploited 
while considering Hartree-Fock or DFT hybrid models where the construction of the Hamiltonian system
is more involved than solving a single Poisson equation.

We note that a practical implementation of the technique can be achieved effectively using the FEAST solver package \cite{feast,fv2}. 
As such, the migration of electronic structure codes making use of SCF mixing schemes, could proceed within two steps:
 (i) integration of the FEAST package as the main  
 linear eigenvalue solver, (ii) reverse-order the SCF-process as it was stated in Section \ref{sec:nlfeast}. Since the
FEAST solver is reverse communication ready, it provides  
the flexibility to build a customized NLFEAST 
with specific computational modules for the problem at hand  (independently of the physical and discretization models).

\begin{figure}[htbp]
\centering
\includegraphics[height=\linewidth,angle=-90]{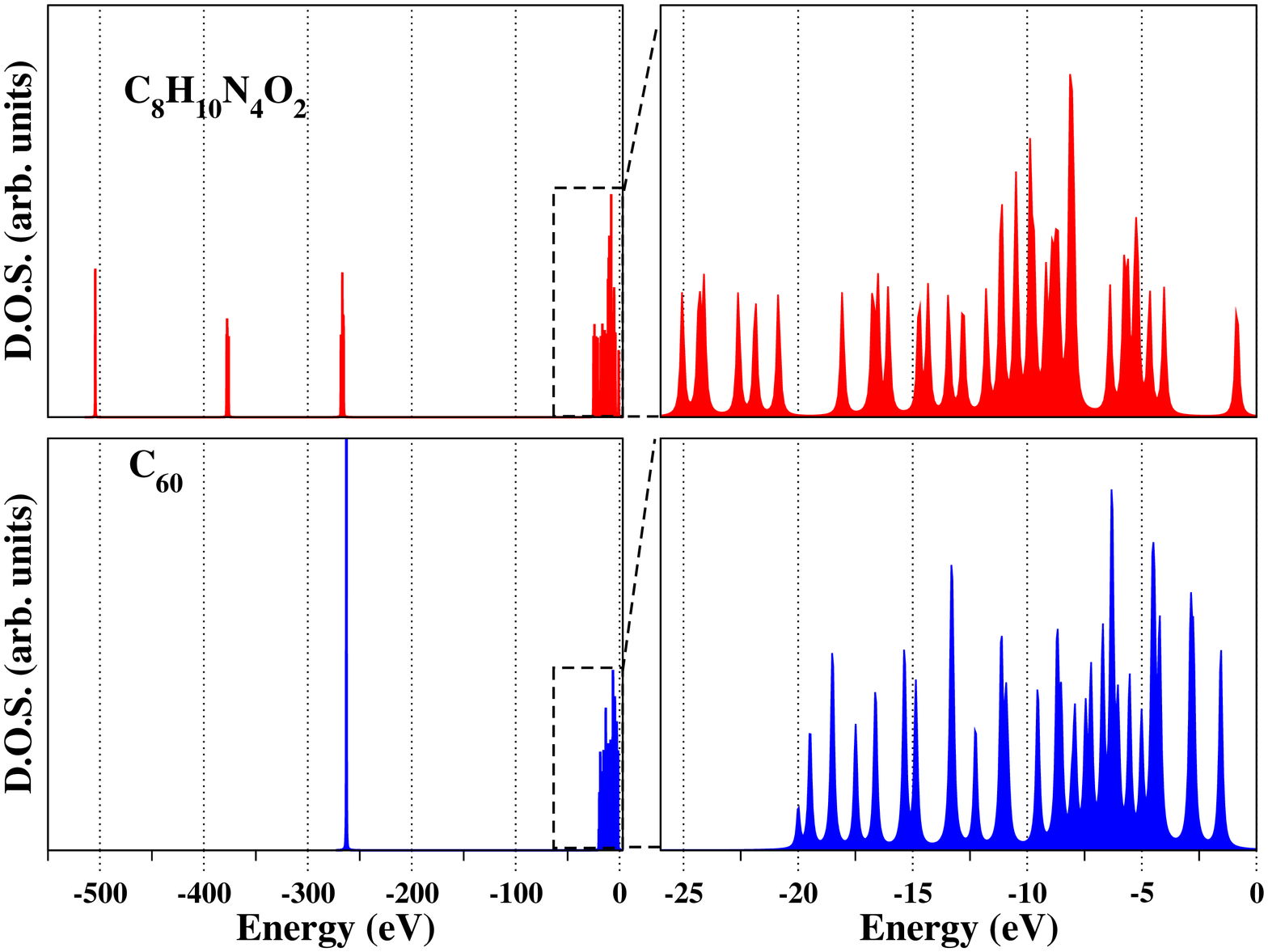}
\caption{\label{fig8} Plots of the density of states for buckminsterfullerene ($\rm C_{60}$) and caffeine ($\rm C_8H_{10}N_4O_2$). 
Our all-electron code is able to capture both the valence states and the core states of each type of atom. From those results, we can clearly identify
the low energy core regions of Oxygen, Nitrogen, and Carbon.}
\end{figure}

In this paper, the applicability of NLFEAST has also been demonstrated for a given search interval which can include several hundred states.
In order to go beyond the current performance capabilities of NLFEAST for addressing much larger molecular systems,
 including nanostructures of current technological interest, explicit parallelism
will become necessary. The parallel treatment of NLFEAST should directly benefit from the intrinsic parallel capability of 
the FEAST algorithm (including the three level of parallelism mentioned in Section \ref{sec:feast}).
For the all-electron model, in particular,  the algorithm can act on different energy ranges (with no overlap for the contour integrations), 
 in order to capture core or valence electrons independently. For example, the results in Figure \ref{fig8} 
clearly illustrate 
 the various core and valence regions within the energy spectrum distribution for a given molecular configuration.

With many search intervals present in the simulation, however,  
one would also need to address the parallel scalability for solving the non-linear reduced system.
Using the linear FEAST within  a traditional SCF process, there is no issue,  since the number of reduced systems is equal to the number of intervals.
Using NLFEAST, however, the situation becomes less straightforward since the construction of the non-linear reduced system  
depends (in principle) on the solutions of the contour integrations for all search intervals. 
The development of new direct or iterative schemes for solving the resulting 
reduced eigenvalue problem that scale linearly with the number of search intervals will need to be addressed in future research.

\begin{acknowledgments}
The authors wish to acknowledge helpful discussions with Prof. Joseph Jerome (The George Washington University), 
Prof. Ahmed Sameh (Purdue University), and Dr. Ping Tak Peter Tang (Intel Corporation). 
This material is supported by NSF under Grant \#ECCS-0846457 and Intel Corporation.
\end{acknowledgments}

%

\end{document}